\documentclass[galaxies,article,accept,pdftex,moreauthors]{Definitions/mdpi} 
\usepackage{pdflscape}	
\usepackage{mathtools}

\firstpage{1} 
\pubvolume{1}
\issuenum{1}
\articlenumber{15}
\pubyear{2023}
\copyrightyear{2023}
\externaleditor{{Academic Editor: Roberta Humphreys}}  
\datereceived{{19 December 2022}}
\dateaccepted{{1 February 2023} }
\datepublished{{}}
\hreflink{https://doi.org/10.3390/galaxies11010026} 

\usepackage{amsmath}
\usepackage{adjustbox}
\usepackage{multirow}
\usepackage{upgreek}
\usepackage{amssymb}
\usepackage{graphicx}
\usepackage{graphics}
\usepackage{multirow}
\usepackage{color}
\usepackage{xcolor}
\usepackage{bm}	
\usepackage{comment}
\usepackage{upgreek}
\usepackage{float}
\usepackage{graphicx}

\def\apj{ApJ}

\def\araa{ARA\&A}
\def\apjl{ApJL}
\def\apjs{ApJS}
\def\aap{A\&A}
\def\mnras{MNRAS}

\def\nat{Nature}
\def\aj{AJ}

\def\J{{$\textit{\textbf{J}}$}}
\def\masyr{\,\mathrm{mas\,yr}^{-1}}

{\newif\ifnotend
\notendtrue
\def\veclist{ABCDEFGHIJKLMNOPQRSTUVWXYZabcdefghijklmnopqrstuvwxyz.}
\def\top#1#2.{#1}
\def\tail#1#2.{#2.}
\loop\expandafter\xdef\csname v\expandafter\top\veclist\endcsname%
{{\noexpand\bf\expandafter\top\veclist}}
\edef\veclist{\expandafter\tail\veclist}
\if\veclist.\notendfalse\fi\ifnotend\repeat}

\def\Gyr{\,\mathrm{Gyr}}

\def\kpc{\,\mathrm{kpc}}

\def\kms{\,\mathrm{km\,s}^{-1}}
\def\masyr{\,\mathrm{mas\,yr}^{-1}}

\def\msun{\,{\rm M}_\odot}

\def\pc{\,\mathrm{pc}}

\def\HI{H\,\textsc{i}}

\def\km2s2{{\rm\,km^2\,s^{-2}}}

\def\ltsima{$\; \buildrel < \over \sim \;$}
\def\simlt{\lower.5ex\hbox{\ltsima}}
\def\gtsima{$\; \buildrel > \over \sim \;$}
\def\simgt{\lower.5ex\hbox{\gtsima}}
\newcommand{\kmskpc}{\ensuremath{\,\mathrm{{km\ s}^{-1}\ {kpc}^{-1}}}}

\Title{The Influence of the Galactic Bar on the Dynamics of \mbox{Globular Clusters}}

\TitleCitation{The Influence of the Galactic Bar on the Dynamics of Globular Clusters}

\Author{Roman Tkachenko $^{1,}$*$^{,\dagger}$\orcidA{}, Vladimir Korchagin $^{1,\dagger}$\orcidB{}, Anna Jmailova $^{2,\dagger}$, Giovanni Carraro $^{3,\dagger}$\orcidC{} and Boris Jmailov $^{4}$ }

\AuthorNames{Roman Tkachenko, Vladimir Korchagin, Anna Jmailova,Giovanni Carraro and Boris Jmailov}
\AuthorCitation{Tkachenko, R.; Korchagin, V.; Jmailova, A.; Carraro, G.; Jmailov, B.}

\address{%
$^{1}$ \quad Institute of Physics, Southern Federal University, Stachki Avenue 124,  344090 Rostov-on-Don, Russia; vkorchagin@sfedu.ru\\
$^{2}$ \quad Physical Faculty, Southern Federal University, Zorge Street 5, 344090 Rostov-on-Don, Russia; zhmailova@sfedu.ru\\
$^{3}$ \quad Dipartimento di Fisica e Astronomia Galileo Galilei, Università degli Studi di Padova, Vicolo Osservatorio 3, I-35122 Padova, Italy; giovanni.carraro@unipd.it\\
$^{4}$ \quad Institute of High Technologies and Piezotechnics, Southern Federal University, 10 Milchakova Street, \mbox{344090 Rostov-on-Don, Russia}; bbzhmaylov@sfedu.ru}

\corres{Correspondence: rtkachenko@sfedu.ru}

\firstnote{These authors contributed equally to this work.} 

\abstract{We make use of recent estimates for the parameters of the Milky Way's halo globular clusters and study the influence of the galactic bar on the dynamics of these clusters by computing their orbits.  We use both an axisymmetric and non-axisymmetric galactic potentials, which include the rotating elongated bar/bulge structure. We account for observational errors both in the positions and in the velocities of the globular clusters and explore the influence of the bar on clusters' evolution. This is contained in the angular momentum--total energy plane, $(L_z, E)$, which is widely exploited as an indicator of the groups of globular clusters that originated from the same accretion event. Particular attention is devoted to the Gaia-Sausage/Enceladus and Pontus structures identified recently as two independent accretion events. Our study shows that it is not possible to identify GSE and Pontus as different merger events.}


\begin{document}


\section{Introduction}

Modern cosmological models predict that the halo of our galaxy was formed by and evolved through the continuous accretion and disruption of star clusters, small satellite galaxies \citep{Amorisco,Chang}, and gas accretion events \citep{Dekel}. 
Our galaxy, the~Milky Way, is a unique and excellent laboratory \citep{Naidu} in which to investigate how this process occurred by studying examples of such accretions, such as the ongoing merging process of the Sagittarius dwarf galaxy \citep{Ibata}; echoes of the past accretion events, e.g.,~ Gaia-Sausage/Enceladus (GSE) at z $\sim$ 2 \citep{Naidu,Belokurov}; or even the first fall beginnings of the Magellanic Clouds \citep{Besla}. As~a result of such mergers, the~halo of Milky Way galaxy, or at least part of it, was built with the products of such accretions as the stellar streams or the globular clusters of extra-galactic~origin.

Our Galaxy consists of a few subsystems---a galactic halo, thick and the thin disks and~an elongated bar/bulge \citep{Port15,Port17}---that have stars with different spatial distributions, velocity dispersions and chemical compositions.  Moreover, even within one subsystem, a few subgroups that have different origins can be often distinguished \citep{Carollo,Helmi,Naidu}. According to \citet{Naidu}, for~example,  most of the stars in the inner halo of the Milky Way seem to be associated with the GSE accretion event. A~comprehensive study of traces of the past mergers is important, therefore, to understand how our Galaxy was formed and how it will evolve in the~future.

The Gaia mission \citep{Perryman,Lindeg,Fabricius,Evanss}, with its accurate data on the coordinates, velocities and metallicities of millions of the halo objects \citep{Wuuu}, opens a new way to study the origin and evolution of the galactic halo, including globular clusters \citep{vas,Baum}, stellar streams \citep{Ib21,Li21} and satellite galaxies \citep{Batt,McC}. 


Based on the last data releases, attempts were made to decipher the nature of globular clusters in the Milky Way using the energy--vertical angular momentum space $(L_z, E)$, where $E$ is a total energy and $L_z$ is the component of the angular momentum of a cluster in the direction perpendicular to the galactic disk. Such attempts were made under the assumption that these objects have values for the energy and z-component of the angular momentum close to the parent accretion event. This idea, proposed by \citet{Helmi2000}, is based on the assumption that the clusters conserve $(L_z, E)$ for billions of years after their accretion onto our galaxy. This picture assumes that the galactic potential was static during billions of years of galactic evolution. In~addition, the~authors did not take into account such effects as the dynamical friction, which significantly changes the energy and the angular momentum of the satellite during its accretion, and~the influence of the Milky Way's elongated bar/bulge potential, which may affect the positions of the clusters on the $(L_z, E)$-plane, especially for those clusters that have small pericentric~radii. 

The aforementioned $(L_z, E)$ classification approach allowed researchers to detect several merger events of the satellite galaxies, such as GSE \citep{Belokurov,Helmi},  Sequoia \citep{Myeong1}, Sagittarius \citep{Ibata},  Kraken \citep{Krui19} and  Pontus \citep{Malhan1,Malhan2}. 
However, \citet{Pagnini}, using  dissipation-less N-body simulations that reproduce the accretion of one or two satellites with their globular cluster population on a Milky Way-type galaxy, it was found that accreted globular clusters do not show dynamical coherence; that is, they do not concentrate in kinematic spaces, if~the mass of the progenitor satellite galaxy is approximately about 10\% of the mass of the Milky Way.
This seems quite contrary to the identification methods presented above. However, for~small accreted satellite galaxies, accretion products (globular clusters) can be confined in a tight range of energies and momentum. In~this case, other mechanisms (in particular galactic bar) may alter the identification of clusters in the same accretion~event.

\citet{Malhan1} used the extended approach and determined the action angles  
\J$ = (J_R, J_\phi ,J_z)$ for known globular clusters, together with their total energies $E$. 
By analysing the distributions of the clusters in $(J_\phi=L_z, E)$, $(J_R, E)$, $(J_z, E)$ and $(J_\phi=L_z, J_R)$, the authors concluded that clusters with similar values of these quantities belong to the same \mbox{accretion event}. 

The Milky Way potential, however, is neither stationary nor axisymmetric. In~the central parts, the~Milky Way potential is influenced by the non-axisymmetric and non-stationary potential of the galactic bar, so the dynamics of the clusters that have small enough pericentric distances will be strongly affected by bar potential after a few passages. In~addition, the~positions and the velocities of the clusters are determined with errors, which also can significantly affect the identification of the clusters in $(L_z,E)$-space. It should be noted here, that we will limit ourselves to studying only $(L_z,E)$-space, because~the calculation of $J_R$ and $J_z$ values in the non-axisymmetric and time-dependent potential with elongated bar/bulge component is not possible. Since the Hamiltonian becomes time-dependent, canonical transformation to action-angle coordinates cannot be obtained \citep{BinTr}. 

In this paper, we study the combined influence of the Milky Way bar potential and  the observational errors to test the identification of GSE and Pontus as two different accretion events \citet{Malhan1}. We recall that the significant influence of the bar on the orbits of objects has been shown in the many previous studies: for globular clusters in \citet{Chemel}; for~stellar streams in \citet{Hattori} and~\citet{Hunt}.
Accordingly, the~layout of the paper is as follows.
Section~\ref{sec:obs} describes available observational data for the globular clusters of the Milky Way halo; 
Section~\ref{sec:model} discusses the model we adopted to study the dynamics of a globular cluster in Milky Way axisymmetric and non-axisymmetric time-dependent potentials. Section~\ref{sec:res} describes the our results of evolution of the globular clusters in $(L_z, E)$ space. Section~\ref{sec:conclusion} summarises the results of our~study.

\section{Observational~Data}\label{sec:obs}
To demonstrate the influence of the galactic bar on the positions of the globular clusters on the ($L_z$, $E$) diagram, we selected two groups of globular clusters that were identified in the literature as the members of two progenitor dwarf galaxies---Pontus and GSE \citep{Malhan1,Malhan2}. The Pontus group consists of seven globular clusters, and~the GSE group consists of 13 clusters listed in Table~\ref{tab1}. In~addition, we include in our considerations the tentative globular cluster NGC~6864/M~75, indicated by  \citet{Malhan1} as a possible candidate for both groups of globular clusters. Table \ref{tab1} provides parameters of the globular clusters, together with their observational errors (inside the parentheses) taken from \citep{vas,Baum}.

\begin{specialtable}
	\widetable
\caption{Heliocentric parameters of globular clusters.}
        \label{tab1}
	\scalebox{0.750}{
		\begin{tabular}{ccccccccc}
			\toprule
			Name & R.A. (deg)& Decl. (deg)&   $D_\odot$ (kpc)& $v_{\rm los}$ ($\kms$)&  \textbf{$\mu^{*}_{\alpha}$} ($\masyr$) & \textbf{$\mu_\delta$} ($\masyr$)\\
            \midrule
            1&2&3&4&5&6&7\\
            \midrule
            \multicolumn{6}{c}{\textbf{Pontus clusters:}} \\
            \midrule
NGC~6341/M~92  	&259.281&	 43.136&	   8.50 (0.07)&	-120.55 (0.27) &	-4.9349 (0.0243)&	-0.6251 (0.0239)\\	
NGC~6779/M~56  	&289.148&	 30.183&	  10.43 (0.14)&	-136.97 (0.45) &	-2.0179 (0.0251)&	 1.6176 (0.0252)\\
NGC~6205/M~13  	&250.422&	 36.460&	   7.42 (0.08)&	-244.90 (0.30) &	-3.1493 (0.0227)&	-2.5735 (0.0231)\\
NGC~7099/M~30  	&325.092&	-23.180&	   8.46 (0.09)&	-185.19 (0.17) &	-0.7374 (0.0246)&	-7.2987 (0.0244)\\
NGC~5286       	&206.612&	-51.374&	  11.10 (0.14)&	  62.38 (0.40) &	 0.1984 (0.0255)&	-0.1533 (0.0253)\\
NGC~288        	& 13.188&	-26.583&	   8.99 (0.09)&	 -44.45 (0.13) &	 4.1641 (0.0241)&	-5.7053 (0.0243)\\
NGC~362        	& 15.809&	-70.849&	   8.83 (0.10)&	 223.12 (0.28) &	 6.6935 (0.0245)&	-2.5354 (0.0242)\\
            \midrule
            \multicolumn{6}{c}{\textbf{tentative cluster:}} \\
            \midrule
            
NGC~6864/M~75  	&301.520&	-21.921&	  20.52 (0.45)&	-189.08 (1.12) &	-0.5975 (0.0262)&	-2.8099 (0.0258)\\            
          \midrule
            \multicolumn{6}{c}{\textbf{Gaia-Sausage/Enceladus clusters:}} \\
            \midrule
NGC~6229       	&251.745&	 47.528&	  30.11 (0.47)&	-137.89 (0.71) &	-1.1706 (0.0263)&	-0.4665 (0.0267)\\
NGC~7492       	&347.111&	-15.611&	  24.39 (0.57)&	-176.70 (0.27) &	 0.7558 (0.0279)&	-2.3200 (0.0276)\\
NGC~6584       	&274.657&	-52.216&	  13.61 (0.17)&	 260.64 (1.58) &	-0.0898 (0.0258)&	-7.2021 (0.0254)\\
NGC~5634       	&217.405&	 -5.976&	  25.96 (0.62)&	 -16.07 (0.60) &	-1.6918 (0.0269)&	-1.4781 (0.0263)\\
IC~1257        	&261.785&	 -7.093&	  26.59 (1.43)&	-137.97 (2.04) &	-1.0069 (0.0400)&	-1.4916 (0.0321)\\
NGC~1851       	& 78.528&	-40.047&	  11.95 (0.13)&	 321.40 (1.55) &	 2.1452 (0.0240)&	-0.6496 (0.0242)\\
NGC~2298       	&102.248&	-36.005&	   9.83 (0.17)&	 147.15 (0.57) &	 3.3195 (0.0255)&	-2.1755 (0.0256)\\
NGC~4147       	&182.526&	 18.543&	  18.54 (0.21)&	 179.35 (0.31) &	-1.7070 (0.0273)&	-2.0896 (0.0274)\\
NGC~1261      	& 48.068&	-55.216&	  16.40 (0.19)&	  71.34 (0.21) &	 1.5957 (0.0249)&	-2.0642 (0.0251)\\
NGC~6981/M~72  	&313.365&	-12.537&	  16.66 (0.18)&	-331.39 (1.47) &	-1.2736 (0.0262)&	-3.3608 (0.0257)\\
NGC~1904/M~79  	& 81.044&	-24.524&	  13.08 (0.18)&	 205.76 (0.20) &	 2.4690 (0.0249)&	-1.5938 (0.0251)\\
NGC~7089/M~2   	&323.363&	 -0.823&	  11.69 (0.11)&	  -3.78 (0.30) &	 3.4346 (0.0247)&	-2.1588 (0.0244)\\
NGC~5904/M~5   	&229.638&	  2.081&	   7.48 (0.06)&	  53.50 (0.25) &	 4.0856 (0.0230)&	-9.8696 (0.0231)\\

			\bottomrule
		\end{tabular}}
	\end{specialtable}

	\noindent{\footnotesize{\textsuperscript{} 1--name of globular cluster;}}
	\noindent{\footnotesize{\textsuperscript{} 2--right ascension (R.A.);}}
	\noindent{\footnotesize{\textsuperscript{} 3--declination (Decl.);}}
	\noindent{\footnotesize{\textsuperscript{} 4--heliocentric distance ($D_\odot$);}}
	\noindent{\footnotesize{\textsuperscript{} 5--spectroscopic line-of-sight velocities ($v_{\rm los}$);}}
	\noindent{\footnotesize{\textsuperscript{} 6, 7--proper motions ($\mu^{*}_{\alpha}=\mu_{\alpha}{\rm cos}\delta,\,\mu_{\delta}$);}} 
	\noindent{\footnotesize{\textsuperscript{}the values in parentheses represent observational errors.}}
	\label{obsdata}
 \noindent

To convert the heliocentric measurements provided in Table~\ref{tab1} into the values in the Galactocentric reference frame, the~\texttt{astropy} package \citep{astro1,astro2,astro3} was used. We adopted for the solar Galactocentric distance the value of $R_{\odot}=8.122$ kpc taken from \citet{Gravity}, and~for the velocity components of the Sun, the values of $ V_{\odot}=(12.9, 245.6, 7.78)$ km/s taken from  \citet{Drim,Reid}. The~displacement of the Sun from the mid-plane of the Galaxy, the value  $Z_{\odot}=20.8$ pc was taken from \citet{Zsun}.

\begin{figure}[H]
\includegraphics[width=14.5 cm]{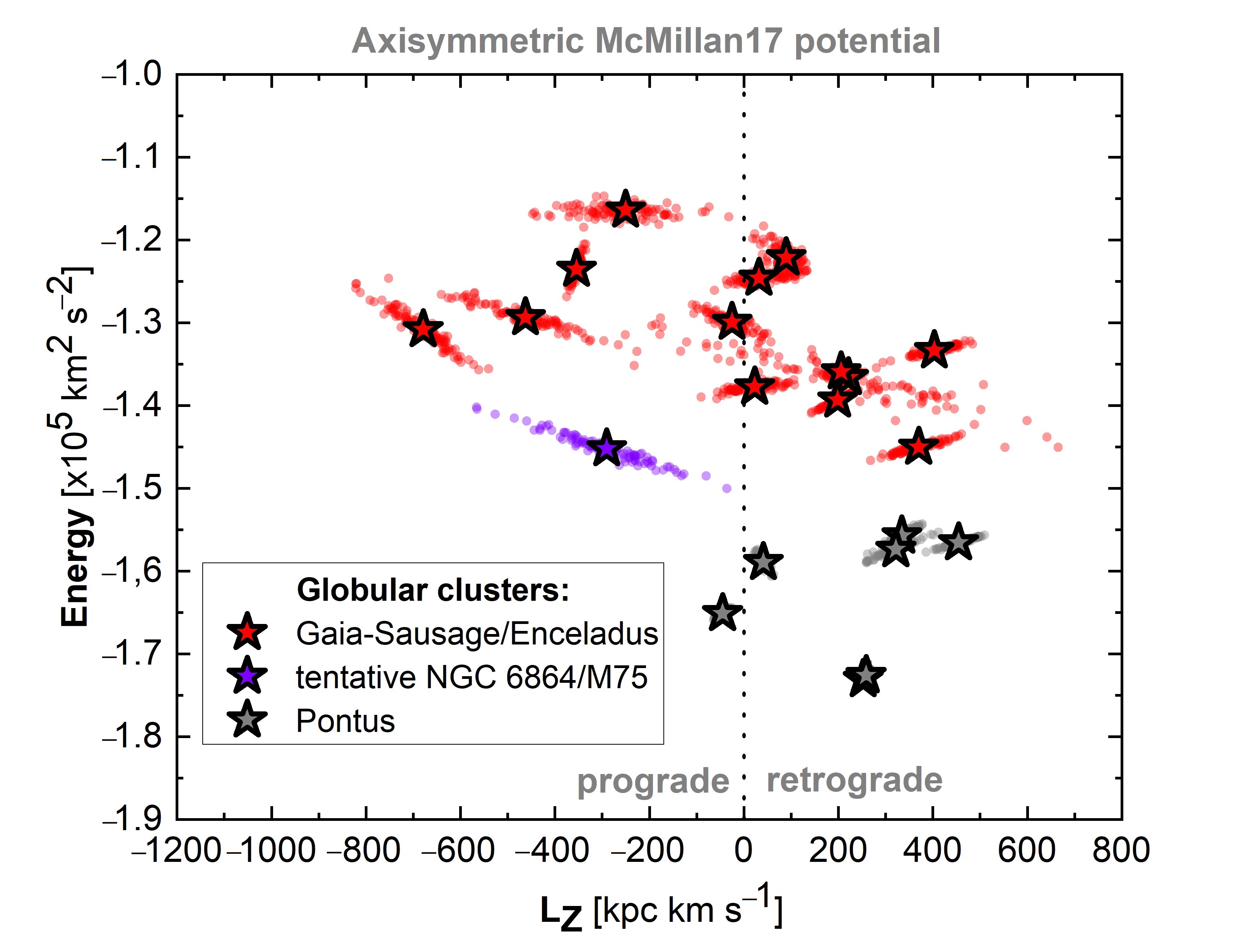}
\caption{Positions of the globular clusters in the $(L_z, E)$-plane. Asterisks mark the mean values of $L_z$ and $E$.
Red---GSE; grey---Pontus globular clusters; purple- the tentative NGC~6864/M~75 cluster. The~dots show one hundred randomly chosen globular cluster
locations consistent with the errors in Table~\ref{tab1}. We have assumed
a normal distribution with the standard deviations set by the errors. \label{fig1}} 
\end{figure}

Figure~\ref{fig1} shows the positions of the globular clusters on the angular momentum--energy plane ($L_z, E$), calculated with the axisymmetric galactic potential \texttt{McMillan17} using data taken from Table~\ref{tab1}.  The~potential model is described in more detail in Section~\ref{sec:model}. Figure~\ref{fig1} shows also the influences of the observational errors on the positions of the globular clusters on the ($L_z, E$)-plane calculated using one hundred randomly chosen coordinates, consistent with the errors in Table~\ref{tab1}(inside the parentheses). We have assumed a normal distribution with the standard deviations set by the~errors.

.


\textls[-15]{As one can readily see in Figure~\ref{fig1}, Pontus objects have z-components of the angular momentum within $L_z\sim[-50 , 500]\kpc\kms$ and energies within $E\sim[-1.75 , -1.55]\times10^5\km2s2$. These values are systematically lower compared to the energies of the GSE group, which have the values of $E\sim[-1.45 , -1.15]\times10^5\km2s2$ and $L_z\sim [-900, 700]\kpc\kms$}. Based on these differences, \citet{Malhan1} came to the conclusion  that these groups of clusters belong to  different merger events. Besides~the values $L_z=J_\phi$ and $E$, \citet{Malhan1} considered the action-angles $J_R$ and $J_z$ as well. In this study, we refrained from this procedure, since the calculation of these values in an non-axisymmetric and time-dependent potential is not possible. We will instead focus  our efforts on the influence of the galactic bar/bulge on evolution of the objects on the $(L_z, E)$-plane. We notice that all objects listed in Table~1 have small pericentric radii ($r_{peri}<3$ kpc), which  makes a strong influence of the bar on these objects. We notice also 
that the globular clusters from Table~\ref{tab1} have both prograde and retrograde orbits, which is caused, probably, by~the internal velocity dispersion of globular clusters that were accreted almost radially on the Milky Way galaxy \citep{Malhan1,Malhan2}.

\section{The~Models}\label{sec:model} 
We conducted simulations of the dynamics of the globular clusters using the astrophysical package \texttt{Galpy} \citep{galpy}. The integration of orbits was performed using a Dormand-Prince (\texttt{dop853}) integrator \citep{SODE}, which is an eighth order Runge--Kutta family method with adaptive timestep. This method was implemented in the package \texttt{Galpy}, and for 1000 orbital periods, it conserved energy at the level of $\Delta E /E=10^{-8}$ in the case of an axisymmetric time-independent potential.

{We integrated the orbits during 4 Gyr forward in time starting from the present epoch, supposing that building of Galaxy was stopped approximately 4 Gyr ago \citep{Griffen}. Therefore, the bar potential remained approximately constant during the last 4 Gyr. The~simulations show that bar stops its evolution after the thickening phase and remains almost unchanged as long as the calculations are performed \citep{Sell,Pfenn}. Therefore, 4 Gyr is enough time to demonstrate that even with the fixed bar potential, the quantities used to pinpoint the past accretion events are not conserved. It also follows from the above that it remains not important to integrate orbits forward or backward in time if we want to demonstrate the influence of the bar.}

\subsection{Galactic Potential~Models}
To study globular cluster dynamics in the Milky Way potential, we used two models. The~first model has an axisymmetric and time-independent potential,
whereas the second one has a non-axisymmetric time-dependent potential with an elongated bar/bulge. 
\subsubsection{Axisymmetric Time-Independent~Potential} 
Following \citet{Malhan1}, first we modelled the Milky Way using the axisymmetric potential described by \citet{McMillan}(\texttt{McMillan17}). The~model included the bulge,
the thick and thin stellar disks, two gaseous disks (\HI\ and $H_2$) and~the
dark matter~halo.

The axisymmetric bulge potential was described by the density profile from \citet{Biss} and can be represented as:
\begin{equation}\label{eq:bulge}
  \rho_\mathrm{b}=\frac{\rho_{0,\mathrm{b}}}{(1+r^\prime/r_0)^\alpha}\;
  \textrm{exp}\left[-\left(r^\prime/r_{\mathrm{cut}}\right)^2\right],
\end{equation}
where  $r^\prime = \sqrt{R^2 + (z/q)^2}$; the~parameters $\alpha$ and $r_0$ were equal to 1.8 and 0.075 kpc, respectively; $r_{\mathrm{cut}}=2.1\kpc$; 
and the axis ratio $q$ was equal to 0.5. The~central volume density of the bulge: $\rho_{0,\mathrm{b}}=9.84\times 10^{10}\,\msun\kpc^{-3}$, which gives a total bulge mass equal to $M_\mathrm{b}=9.23\times 10^9\,\msun$.

The Milky Way disk consists of two subsystems, the~thin and the thick \citep{Gilm} exponential disks both described by density profiles with the form:
\begin{equation}\label{eq:disc}
  \rho_\mathrm{d}(R,z)=\frac{\Sigma_{0}}{2z_\mathrm{d}}\;\textrm{exp}\left(-\frac{\mid
      z\mid}{z_\mathrm{d}}-\frac{R}{R_\mathrm{d}}\right).
\end{equation}  
Here, $z_\mathrm{d}$, $R_\mathrm{d}$ and $\Sigma_{0}$ are the vertical scale height,  the~radial scale length and~central surface density, respectively. The~vertical scale heights of the disks are fixed to the values  $z_{\mathrm{d},\mathrm{thin}}=300\pc$ and 
$z_{\mathrm{d},\mathrm{thick}}=900\pc$, and~the scale lengths for the thin and thick
disks are assumed to be $2.5\kpc$ and $3.02\kpc$, respectively.  With~the central surface densities of the thin and of the thick disks taken to be $\Sigma_{0,\mathrm{d},\mathrm{thin}}=896 \msun\pc^{-2}$ and $\Sigma_{0,\mathrm{d},\mathrm{thick}}=183 \msun\pc^{-2}$, the~total masses of the thin and of thick disks can be estimated as $M_\mathrm{d}=2\pi\Sigma_{0}R_\mathrm{d}^2$ and equal to $3.518\times 10^{10}\,\msun$ and $1.048\times 10^{10}\,\msun$, respectively.
A possible central 'hole' in stellar density is not~considered.

The \texttt{McMillan17} potential also includes two gaseous disks: 
the atomic ($\HI$) one and the molecular ($H_2$) one. The~density distributions in both gaseous disks are described by this equation:
\begin{equation}\label{eq:gasdisc}
  \rho_\mathrm{d}(R,z)=\frac{\Sigma_{0}}{4z_\mathrm{d}}\;
  \textrm{exp}\left(-\frac{R_{\rm m}}{R}-\frac{R}{R_\mathrm{d}}\right)\;
  {\rm sech}^2(z/2z_d).
\end{equation}
Here, $R_{\rm m}$ gives the scale length of central gaseous hole. For~the $\HI$ disk, the following parameters are adopted: $R_\mathrm{d,\HI}= 7 \kpc$, $z_{\mathrm{d},\mathrm{\HI}}=85\pc$, $\Sigma_{0,\mathrm{d},\mathrm{\HI}}=53.1 \msun\pc^{-2}$, $R_{\rm m,\HI}= 4 \kpc$. For~the $H_2$ gaseous disk, the accepted values are $R_\mathrm{d,H_2}= 1.5 \kpc$, $z_{\mathrm{d},\mathrm{H_2}}=45\pc$, $\Sigma_{0,\mathrm{d},\mathrm{H_2}}=2180 \msun\pc^{-2}$ and~$R_{\rm m,H_2}= 12 \kpc$.

The potential of the dark matter halo is described by the simple density profile given by this equation:
\begin{equation} \label{eq:halo} \rho_{\mathrm{h}} =
  \frac{\rho_{0,\mathrm{h}}}{x^\gamma\,(1+x)^{3-\gamma}},
\end{equation}
where $x=r/r_{\mathrm{h}}$, and~$r_{\rm h}=19.6 \kpc$ is the scale length of the density distribution, and~$\rho_{0,\mathrm{b}}=0.00854\times 10^{10}\,\msun\kpc^{-3}$ is a characteristic scaling density of the halo. With~the parameter $\gamma=1$, Equation~(\ref{eq:halo}) becomes a classic NFW \citep{NFW} density profile. The mass of the dark-matter halo (within the sphere of 200 kpc) is equal to $M_{200}=1.216\times 10^{12}\,\msun$.

All parameter values came from \texttt{McMillan17} \citep{McMillan}. To~integrate the equations of motion, it is necessary to calculate the potentials using known density distributions. The~NFW potential has  analytical representation. The~potential of the bulge was calculated using the basis-function expansion of the self-consistent-field method of \citet{Hernq}. Potentials for disks were calculated using a modified technique  of \citet{Hernq}, which was presented in \citet{Kuij} and implemented in the\texttt{Galpy} package \citep{galpy}.

\subsubsection{Non-Axisymmetric Time-Dependent Potential with Elongated Bar/Bulge} 

In this model, we replaced the axisymmetric bulge with the non-axisymmetric elongated bar/bulge~component. 

Following \citet{Chemel} and \citet{Yeh}, we
approximate the bar/bulge by a rotating ellipsoid with
a density distribution given by a \texttt{Ferrers} model \citep{galpy}:
\begin{align}
\label{bar_density}
\rho(x,y,z) = \begin{cases}
\rho_c (1-m^2)^2&\text{, if~m} < 1, \\
0 &\text{, if~m} > 1,
\end{cases}
\end{align}
where $m^{2}=\dfrac{x^2}{a^2}+\dfrac{y^2}{b^2}+\dfrac{z^2}{c^2}$. For~the bar semi-axes, we chose the values $a = 5$ kpc, $b = 2$ kpc and~$c = 1$ kpc, which gives a good approximation for the
elongated bar/bulge  observed in the Milky Way (see, for~example, Model \texttt{M85} in \citet{Port15,Port17}). The~central density of the bar in this model is determined by the expression $\rho_c=\dfrac{105}{32\pi}\dfrac{GM_{b}}{abc}$, where $M_{b}$ is the mass of the bar, for~which we adopted the value $M_\mathrm{b}=1.88\times 10^{10}\,\msun$ from \citet{Port17} (also similar to \citet{Kent}).  \citet{Zhao} built a self-consistent model of the bar/bulge and found the total mass of the bar taken to be equal $2\times 10^{10}\,\msun$ \citep{Bland}. 
\citet{Kipper} estimated a smaller value of $\sim1.6\times 10^{10}\,\msun$

The gravitational potential produced by the density in Equation~(\ref{bar_density}) can be \citep{Pf,Yeh}:
\begin{equation} 
\Phi = -\pi G abc \dfrac{\rho_c}{n+1} \int_{\lambda}^{\infty}\dfrac{du}{\Delta(u)}(1-\zeta^2(u))^3 \text{,} 
\end{equation}
\begin{equation}
\text{where  }  \zeta^2(u)=\dfrac{x^2}{a^2+u}+\dfrac{y^2}{b^2+u}+\dfrac{z^2}{c^2+u} \text{, }
 \end{equation}
\begin{equation}
 \text{ and } \Delta^2(u)=(a^2+u)(b^2+u)(c^2+u),
\end{equation} 
where n is integer number, which is equal to two, according to the degree in Equation~(\ref{bar_density}). $\lambda$ is the unique positive solution of $\zeta^2(\lambda)=1$, outside the bar ($\zeta \geq 1$). Inside the bar, $\lambda=0$ (for the more details see \citet{Pf}).

\begin{figure}[H]
\includegraphics[width=13.5 cm]{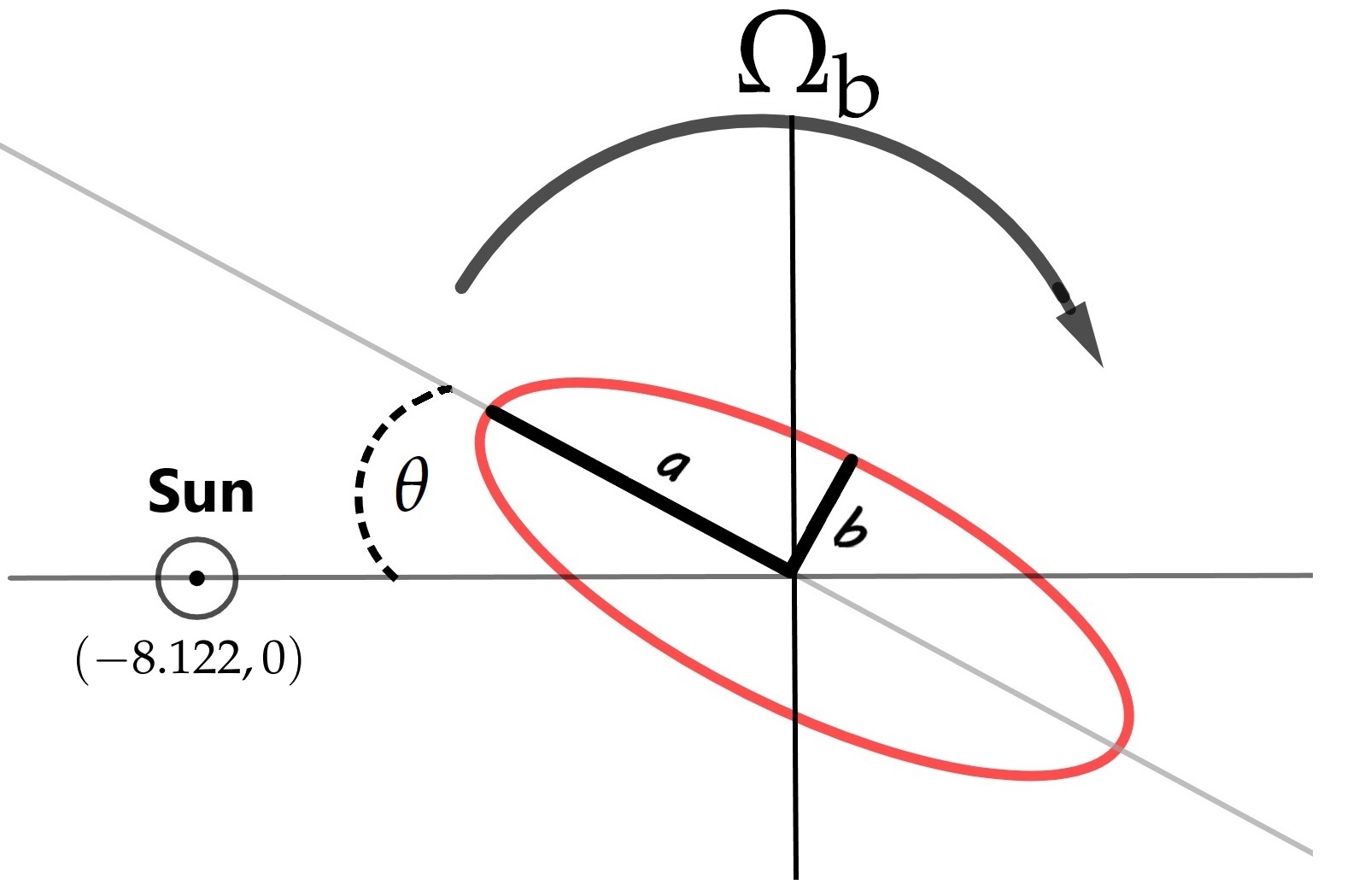}
\caption{Schematic representation of the position of the Milky Way bar relative to the Sun. The angle $\theta$ between the Sun and the bar is $28^{\circ}$. The~bar has an angular rotation velocity $\Omega_{\rm b}$ and rotates clockwise, so $\Omega_{\rm b}$ has a negative~sign.
\label{figbar}} 

\end{figure}

The Milky Way bar rotates with  constant angular velocity, and~the major axis is tilted with respect to the Sun, as shown in Figure~\ref{figbar}.

A number of studies estimated the bar orientation relative to the Sun (\citep{Port17,Bland,Port15,Sanders,Wegg}) and its angular velocity (\citep{Debat,Sanders,Clarke,Shen,Sormani}). \citet{Debat} were the first to estimated the angular velocity of the galactic bar: $\Omega_{\rm b}= 59\pm5 \kmskpc$. Recent data that used Gaia proper motion measurements allow one to estimate the pattern speed of the bar more accurately, although~the spread in the measurements remains considerable. \citet{Sanders} using the VVV Infrared Astrometric Catalogue (VIRAC) and Gaia DR2 proper motions obtained the value of $\Omega_{\rm b}= 41\pm3 \kmskpc$. \citet{Clarke} obtained  $\Omega_{\rm b}= 37.5 \kmskpc$ \citep{Shen}. \citet{Sormani}, using hydrodynamic gas simulations in barred Milky Way potential, found that the angular velocity of the bar is $\Omega_{\rm b}= 40 \kmskpc$. We chose in our non-axisymmetric model the value of $\Omega_{\rm b}= 40 \kmskpc$, which is in a good agreement with \citet{Port15,Port17} and \citet{Bovy19}.

Recent measurements of the bar orientation  $\theta$ relative to the Sun lie in the range of  $23^{\circ}-33^{\circ}$ \citep{Port17,Bland,Port15,Sanders}. We used in our simulations the value $\theta=28^{\circ}$, which is in a good agreement with the most recent estimates \citep{Wegg,Port17,Bland}.

\subsection{Rotation Curves: Comparison of Axisymmetric and Non-Axisymmetric~Models}
\begin{figure}[H]
\includegraphics[width=14.0 cm]{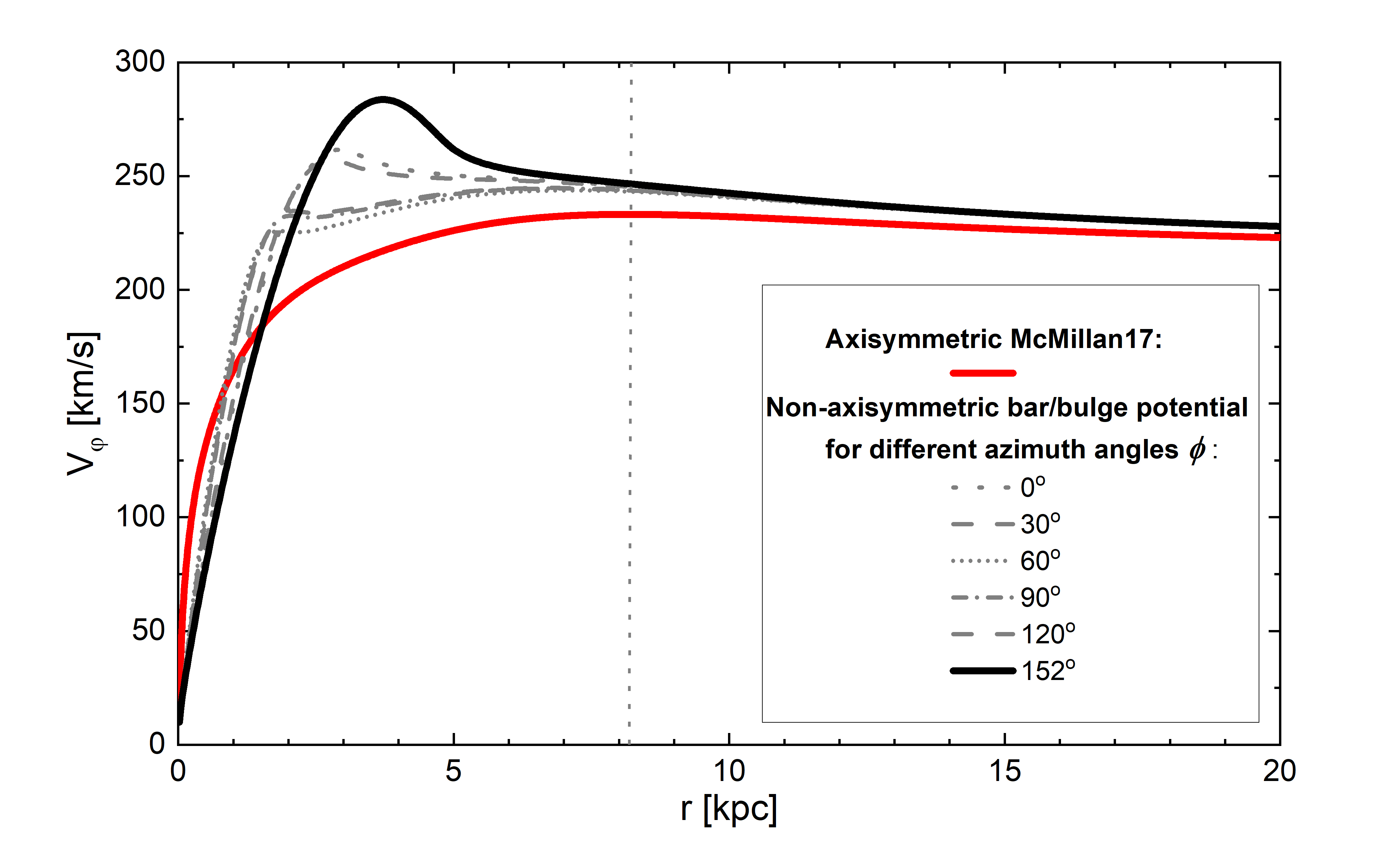}
\caption{Rotation curves for axisymmetric \texttt{McMillan17} (red line) and for a non-axisymmetric model with the bar for different  orientations (black lines). Angle of $152^{\circ}$ corresponds to the direction along the major axis of the bar.\label{figRC}}
\end{figure}

Figure~\ref{figRC} shows the Milky Way rotation curves in our barred and~non-barred models. The~model with the bar has systematically higher values of the rotational velocity compared to the axisymmetric model \texttt{McMillan17} due to the fact that in the model with the bar, we replaced the axisymmetric bulge in \texttt{McMillan17} with an elongated bar/bulge whose mass is about two times larger than the mass of the bulge in the \texttt{McMillan17} model. Rotation curves in Figure~\ref{figRC} were derived (in the plane z=0) from the equation $v(r)=(r\,d\Phi /dr)^{1/2}$, where $\Phi$ is the total potential.

\section{Results and~Discussion}\label{sec:res}
We discuss in this section the dynamics of the halo globular clusters and the influence of the galactic bar on their positions in the ($L_z, E$)-plane.

\begin{figure}[H]
\includegraphics[width=11.0 cm]{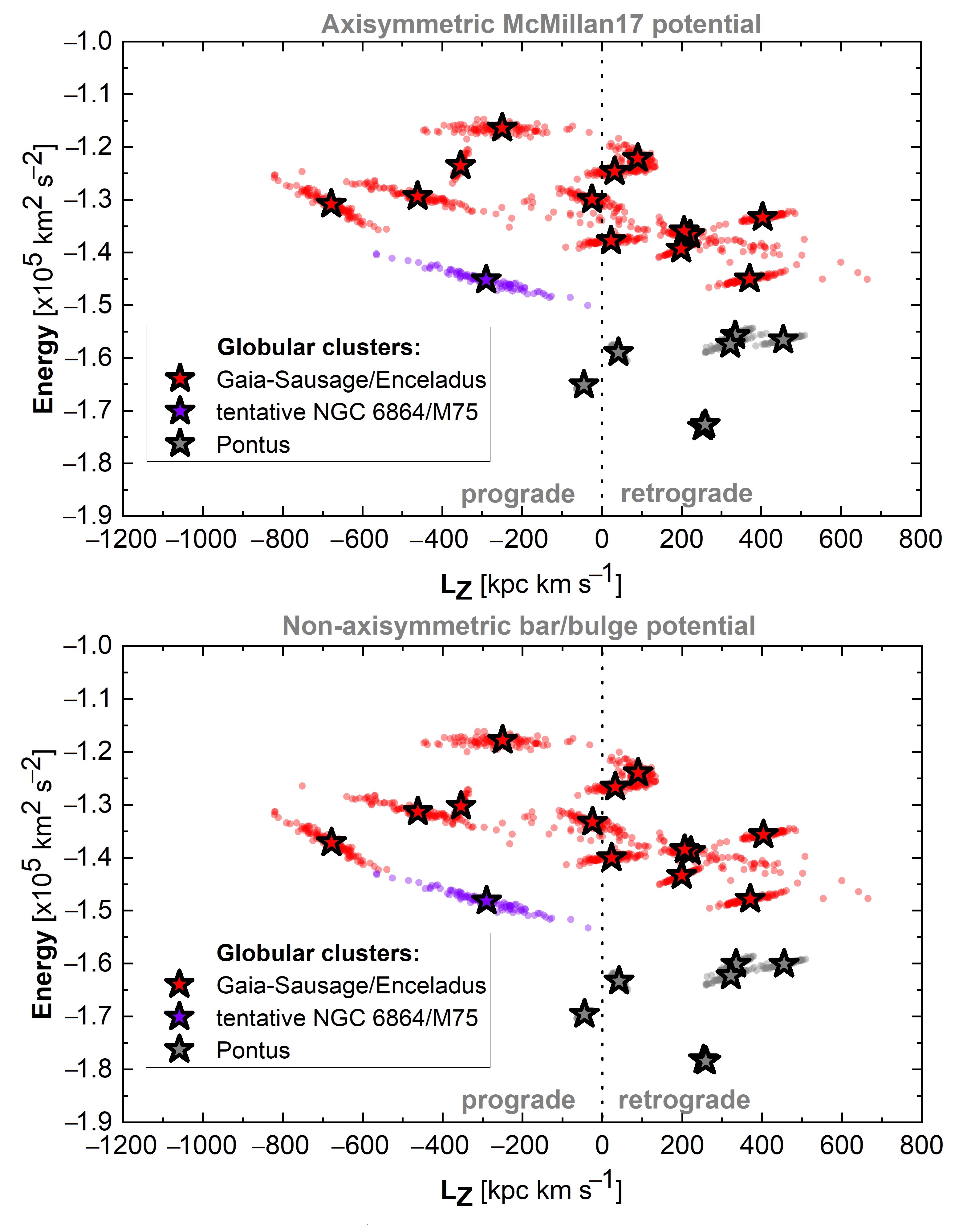}
\caption{Positions of the globular clusters on the $(L_z, E)$-plane for two galactic potential models, the axisymmetric model of \texttt{McMillan17} (top panel) and a non-axisymmetric model (bottom panel). Asterisks mark the mean values of $L_z$ and $E$ on the~plane.\label{Fig4}} 
\end{figure}

Figure~\ref{Fig4} shows the ($L_z, E$)-positions of the globular clusters calculated using data taken from Table~\ref{tab1} for barred and non-barred galactic potential models. Figure~\ref{Fig4} also shows one hundred alternative positions of each cluster considering observational errors.
As one can readily see, the~clusters have similar values of $L_z$ and $E$ with minor differences. Note, however, that while in the axisymmetric and time-independent potential, the values of $L_z$ and $E$ are conserved,  in~time-dependent non-axisymmetric potential, these values are not conserved in time. We will discuss in detail the non-conservation of $L_z$ and $E$ below.

First, let us consider the influence of the galactic bar on the positions of the clusters on ($L_z, E$)-plane using mean values of their parameters (asterisks in Figure~\ref{Fig4}) in time-dependent non-axisymmetric~potential.

\begin{figure}[H]
\includegraphics[width=12.5 cm]{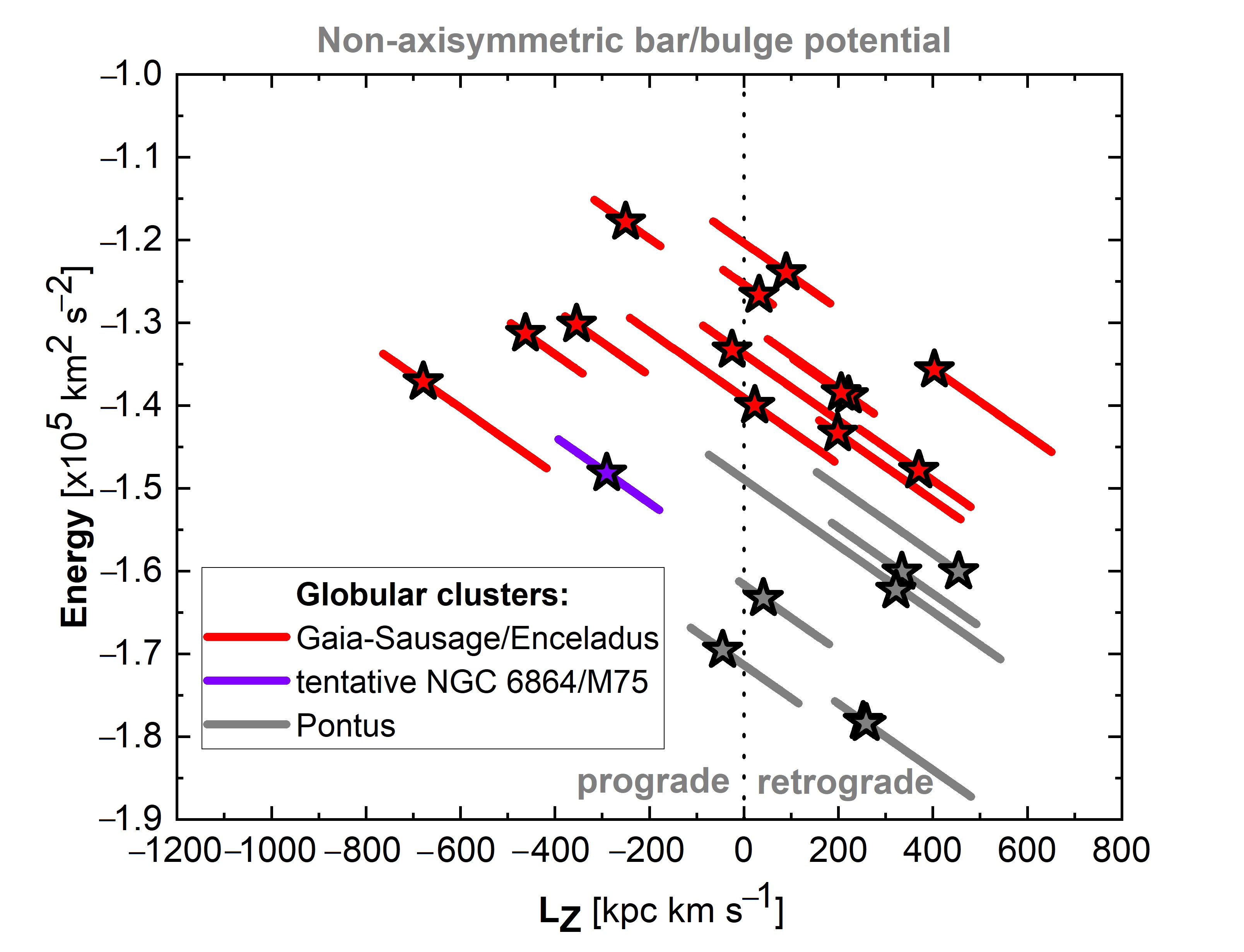}
\caption{Evolution of the globular clusters on the $(L_z, E)$-plane for non-axisymmetric potential with the rotating bar. Asterisks mark the initial positions of the clusters in the $L_z,E$-plane (red colour---GSE objects; grey---Pontus objects; purple---tentative NGC~6864/M~75 cluster). The~lines show trajectories along which the globular clusters~oscillate.\label{fig5}} 
\end{figure}

To do this, we integrated the orbits of each cluster during 4$\Gyr$ in the non-axisymmetric potential that includes the rotating bar/bulge. The~evolution of each globular cluster on the ($L_z, E$)-diagram is shown in Figure~\ref{fig5}. The~initial positions of clusters in ($L_z, E$) space are indicated in Figure~\ref{fig5} with asterisks. As~can be seen in the Figure, in~the non-axisymmetric potential of the bar, rotating with the constant angular velocity $\Omega_{\rm b}$, the~clusters oscillate along a straight line that has the same slope for each cluster. This is due to the fact that in the non-axisymmetric time-dependent potential, the~z-component of the angular momentum $L_z$ and the total energy $E$ of the object are not conserved. 
Instead, the~Jacobi integral described by the equation:
\begin{equation}\label{eq:Jacob}
    E_{\rm J} = E - \Omega_{\rm b} L_z .
\end{equation}
is conserved for each cluster (see, e.g.,~\citet{Hattori}).
Equation~(\ref{eq:Jacob}) shows that in case of constant angular velocity of the bar, the total energy $E$ is a linear function of $L_z$, resulting in the clusters oscillating in Figure~\ref{fig5} along parallel~lines.   
 
As can be seen in Figure~\ref{fig5}, the~positions of GSE and Pontus clusters are significantly affected by the Milky Way bar, which blur their positions on the $(L_z, E)$ diagram over time. Energy variation for clusters is up to  15\%, and~variation of the angular momentum is more than 100\%. It should also be noted that some Pontus objects reach energy values of the GSE~group. 

We then took into account the errors in observational data of coordinates and proper motions, leading to the uncertainties in the initial galactocentric coordinates and the velocities of the the globular clusters. To~do this, we randomly generated one hundred sets of the coordinates in the 6D phase space for each globular cluster. They were drawn assuming Gaussian errors in the measured coordinates and velocities, and~then we calculated the positions of the clusters on the $(L_z, E)$-plane.

\begin{figure}[H]
\includegraphics[width=14.0 cm]{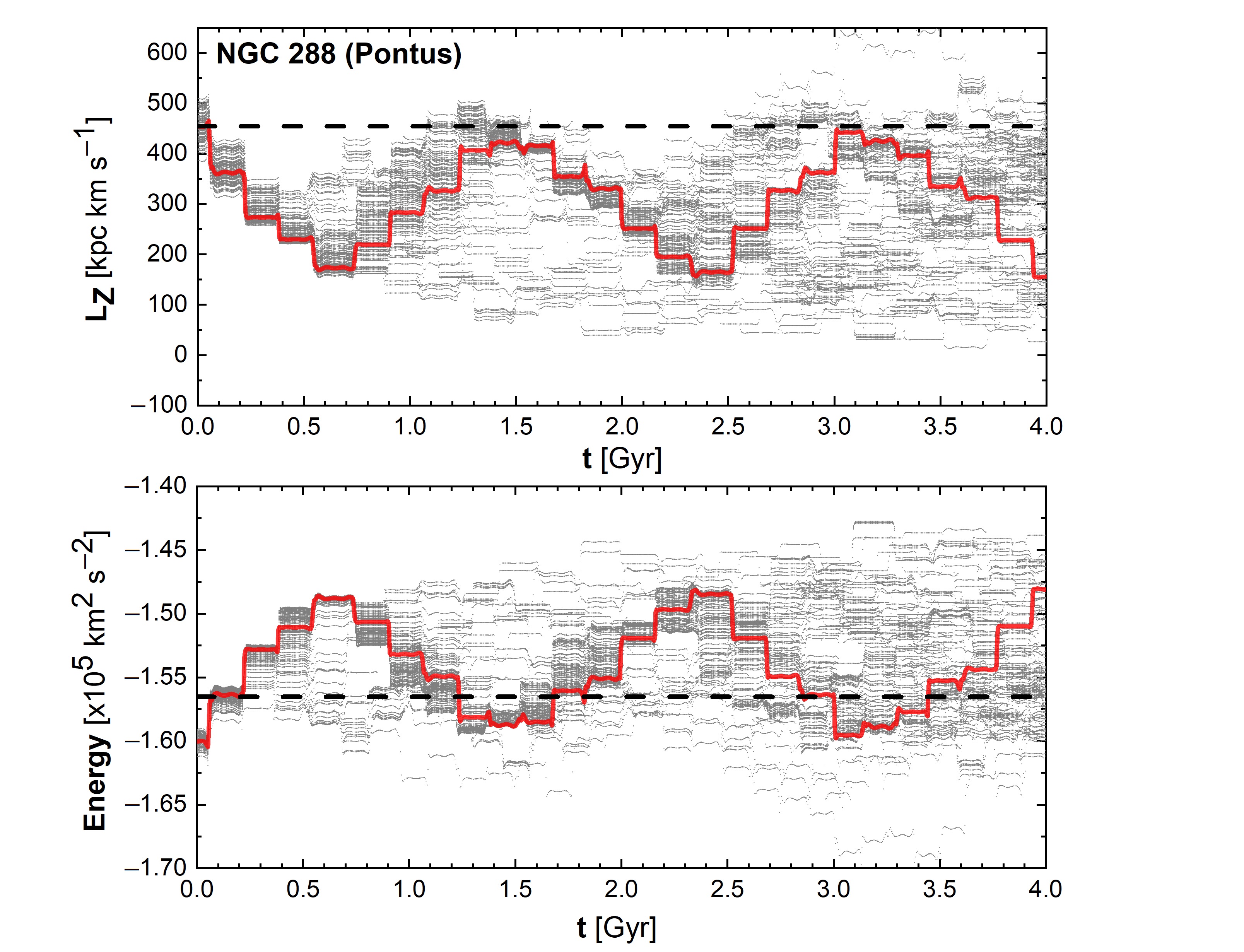}
\caption{Evolution of the total energy $E$ (bottom frame) and the z-component of angular momentum $L_z$ (upper frame) for the globular cluster NGC~288 evolving in the non-axisymmetric rotating potential of the bar. The~red line shows the evolution for the mean value of initial conditions. The~grey lines are one hundred representations of the orbits taking into account the observational errors in coordinates and proper motions. The~black dashed line shows the constant angular momentum and energy in the axisymmetric \texttt{McMillan17} potential.Time $t=0$ means the start time of the simulation and corresponds to the present~time. \label{fig6}}
\end{figure}

Before discussing the evolution of the globular clusters on the $(L_z, E)$-plane, a~closer look at the dynamics of one globular cluster in the Pontus group---NGC~288---is necessary.  Figure~\ref{fig6} shows time dependence of the angular momentum $L_z$ and the total energy $E$ for the mean values of the
initial conditions (red line), while~also taking into account observational errors (100 grey dotted lines).
As one can see, the~angular momentum and the energy of the cluster are significantly affected by the time-dependent potential of the rotating bar.
The values oscillate with time and correlate with each other according to Equation~(\ref{eq:Jacob}). The stepwise changes in momentum and energy are due to the moments of time when the cluster comes close to the bar and passes through it. Areas with almost constant values refer to the moments in time when the cluster was moving far from the bar.
Observational errors spread the values of the angular momentum and the energy of the cluster. After only 1--2 Gyr of evolution, the~range of values consistent with the observational error is quite large. The~spread of the values of the angular momentum is within $L_z\sim[-0 , 600]\kpc\kms$, and total energy of the cluster gets the values $E\sim[-1.70, -1.42]\times10^5\km2s2$. For~comparison, the~black dotted lines in Figure \ref{fig6} showed the constant values of $L_z$ and $E$ in the time-independent axisymmetric potential of \texttt{McMillan17}.

\begin{figure}[H]
\widefigure{
	\includegraphics[width=0.70\textwidth]{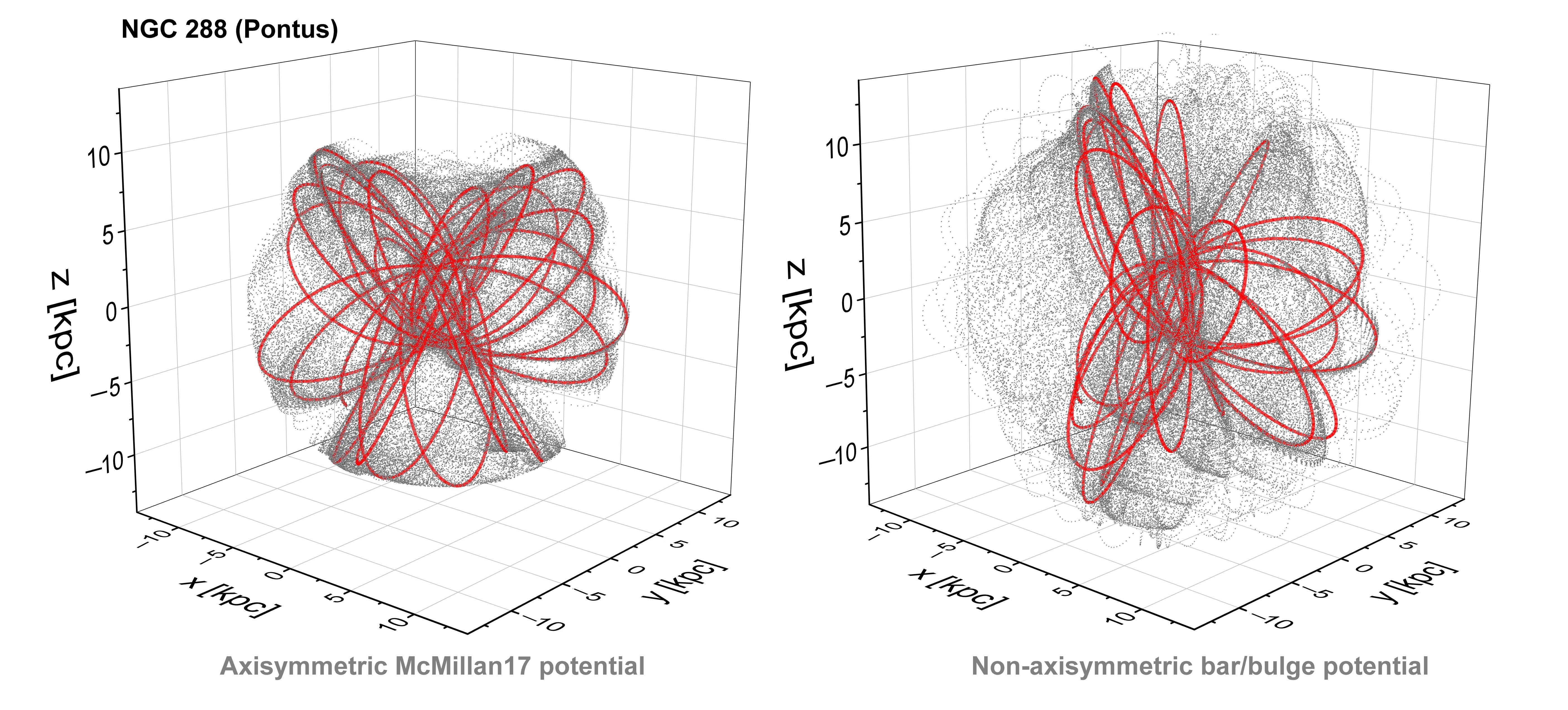} }
	\caption{Three-dimensional orbits of globular cluster NGC~288 in the axisymmetric potential \texttt{McMillan17}(left panel) and in the non-axisymmetric rotating bar/bulge potential (right panel). The~red lines show the orbits for the mean initial conditions. The~grey lines show one hundred orbits consistent with the observational errors in these initial~conditions.\label{fig7}} 
\end{figure}

\begin{figure}[H]
\widefigure{
	\includegraphics[width=0.70\textwidth]{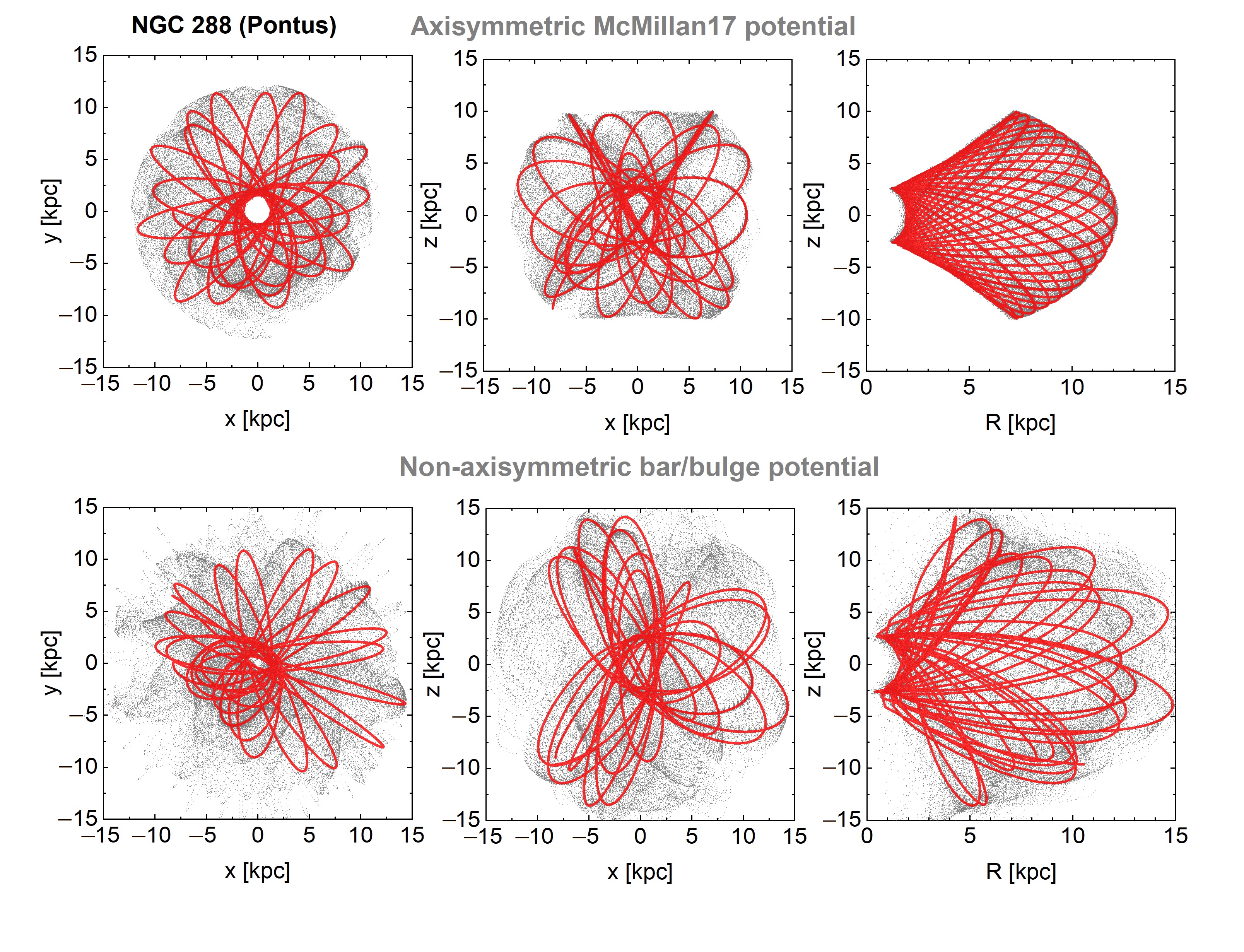} }
	\caption{Projections of the cluster orbits onto the galactocentric coordinate planes for the axisymmetric potential \texttt{McMillan17}(top frames) and for the non-axisymmetric potential of the rotating bar (bottom frames). The~red lines show the orbit for the mean initial conditions, whereas the grey lines show one hundred orbits consistent with the observational errors in these initial~conditions.\label{fig8}}
\end{figure}

Figure~\ref{fig7} shows the three-dimensional orbits of the globular cluster NGC~288 in the axisymmetric potential \texttt{McMillan17} and in the non-axisymmetric potential with the rotating bar. 
Figure~\ref{fig8} shows the orbits of the cluster projected onto the three galactocentric planes. The~mean orbits are shown as red lines, whereas one hundred alternative orbits consistent with the observational errors of these initial conditions are shown in grey.
We can see that the bar significantly affects the dynamics of the globular cluster: the bar significantly randomises the orbits and destroys orbital boxes (diagram R vs. z in Figure~\ref{fig8}), confirming the significant influence of the bar stressed by \citet{Chemel} and \citet{Hattori}.  
The cluster NGC~288 has a small pericentric distance similar to the other clusters considered in this paper. For~the clusters that have larger pericenters, the influence of the bar will obviously decrease due to the increase in the distance of the symmetry of the equipotential surfaces of the~bar. 

\nointerlineskip
\unskip

\nointerlineskip

McMillan17 (left panel) and the non-axisymmetric rotating bar/bulge potential (right panel): the~red lines show the orbits for the mean initial conditions. The~grey lines show one hundred orbits consistent with the observational errors in these initial~conditions.

Finally, let us consider the joint influence of the bar and of the observational errors on the evolution of the clusters in the $(L_z, E)$-plane. To~do this, we took one hundred representations of the orbits using the errors in Table~\ref{tab1} and represented by dots in Figure \ref{Fig4}.

\begin{figure}[H]
\includegraphics[width=14.5 cm]{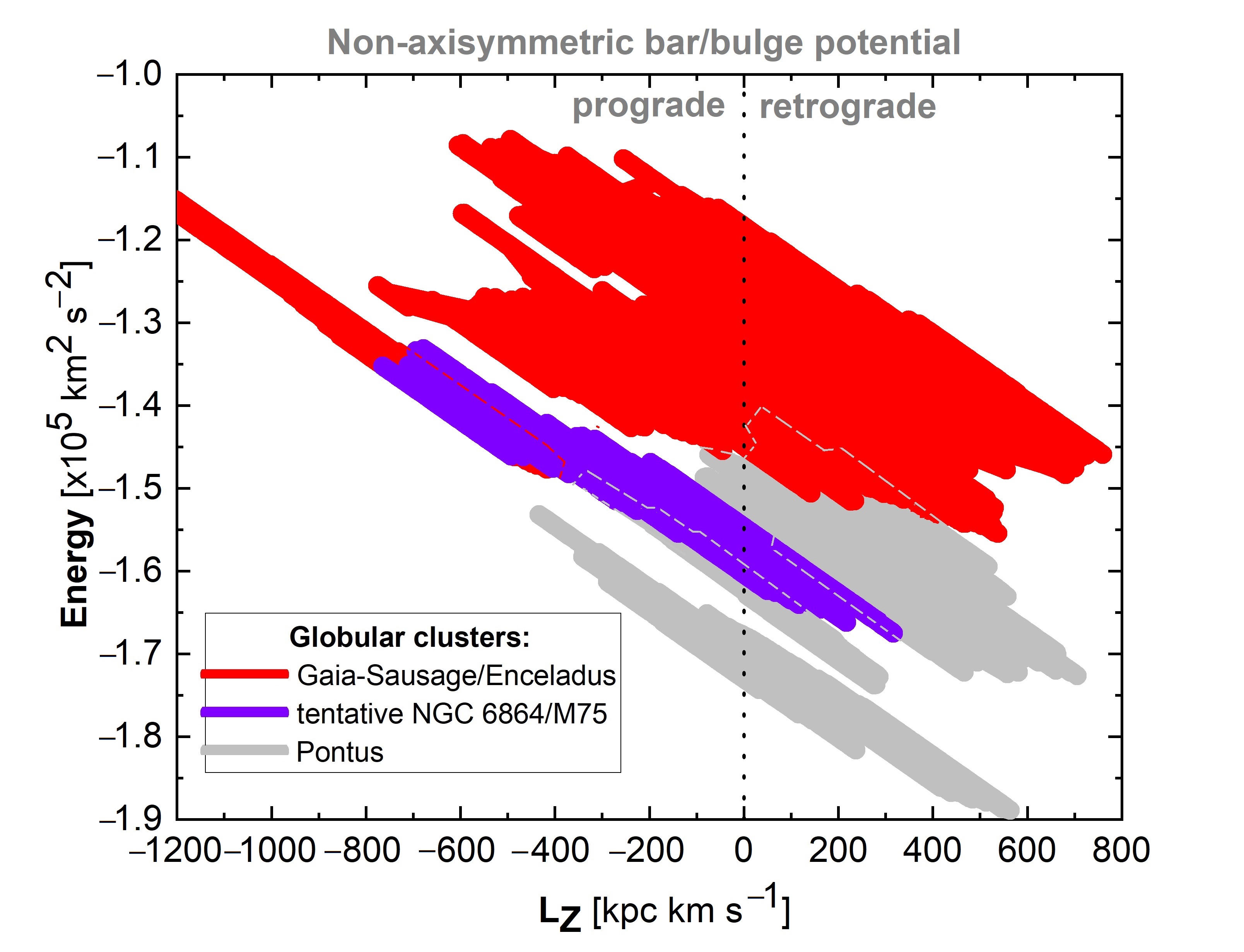}
\caption{Evolution of the globular clusters from Table~\ref{tab1} during 4 Gyr after the beginning of simulations, taking into account the non-axisymmetric potential of the rotating bar and
the observational errors of the initial conditions.}
\label{fig9}
\end{figure}

This is illustrated in Figure~\ref{fig9}, which shows the evolution of the clusters---represented by one hundred dots for each clusters in Figure \ref{Fig4}---in the non-stationary potential of the rotating bar, and~under the influence of the uncertainties in the initial kinematical data of the clusters. It is difficult to conclude that these clusters belong to two different accretion events, namely, Pontus and GSE, as was suggested by \citet{Malhan1}.
The "tentative" cluster NGC~6864 overlaps with both groups. Thus, it cannot be concluded that GSE and Pontus clusters belong to different accretion events and so to different progenitor galaxies. The~same can be said about the origin of the tentative object NGC~6864.
It is also impossible to make firm conclusion on which group the tentative object NGC~6864 belongs~to. 

In more general terms, we conclude that it is difficult to identify accretion events using objects that have close passages around the galactic centre. For~those objects, the bar significantly affects the values of $E$ and $L_z$. One should also take into account that during accretion of a dwarf galaxy, dynamical friction constantly changes its energy and momentum, making the identification of the original accretion events even more~difficult.

Finally, we did not take into account the influence of other important physical phenomena that can also affect the positions of the clusters on the $L_z,E$-diagram such as the change of the galactic potential due to galaxy's accretion and secular evolution, the~change of the angular velocity of bar during its evolution, the~influence of the spiral structure,  and~the interaction of globular clusters with satellite galaxies on evolution of the globular clusters. 
We will attempt to include the effect of these phenomena in the~future.

\section{Summary}\label{sec:conclusion}

In this study, we concluded that the bar significantly affects the dynamics of the globular clusters with small enough pericentric distances. We found that the bar potential renders the orbits of such clusters quite chaotic, confirming the results obtained by \citet{Chemel}. 
If one takes into account the bar potential and the observational errors in the heliocentric coordinates and in the proper motions of the clusters, it is evident that the positions of the clusters on $L_z, E$-diagram are not tight anymore, making it impossible to identify the cluster's association with a particular accretion event or progenitor galaxy. This result is especially true for those clusters that pass close enough to the galactic centre and have the perihelion less than about 3 kpc, i.e.,~where the influence of the bar is stronger. 
To stress it yet again, in~this work we provided strong evidence that a proper accounting of the influence of the bar, together with observational errors, is mandatory. We have applied this to two particular accretion events---the GSE and the Pontus---and clearly showed that it is impossible to discriminate among different parent merger~events.

 
\vspace{6pt} 


\authorcontributions{Conceptualisation, V.K., G.C. and R.T.; methodology, V.K., G.C. and R.T.;
software, R.T., A.J. and B.J.; writing—original draft preparation, R.T., V.K. and G.C.; writing—review and
editing, R.T., V.K., A.J, G.C. and B.J. All authors have read and agreed to the
published version of the~manuscript.}

\funding{RT thanks the Foundation for the Advancement of Theoretical Physics and Mathematics 'BASIS' for financial support at \url{https://basis-foundation.ru}, accessed on 1 July 2022.} 

\institutionalreview{Not applicable.}

\informedconsent{Not applicable.}

\dataavailability{All data used in this paper were taken from the open sources, and the references are given.} 

\acknowledgments{We thank the anonymous reviewers for the careful reading of the article and valuable~comments.}

\conflictsofinterest{The authors declare no conflict of~interests.} 


\abbreviations{Abbreviations}{
The following abbreviations are used in this manuscript:\\

\noindent 
\begin{tabular}{@{}ll}
GSE & Gaia-Sausage/Enceladus\\
NFW & Navarro--Frenk--White  \\
\end{tabular}
}


\end{paracol}


\reftitle{References}

\end{document}